# High-Quality Ge-Doped (010) β-$Ga_2O_3$ Homoepitaxial Films Grown by Low-Pressure CVD: Structural, Electrical, and Schottky Diode Characteristics

Ahmed Ibreljic[1], Saleh Ahmed Khan[1], Sourav Sarker[1], Stephen Margiotta[1], Stephen Lam[2], Anhar Bhuiyan[1, a)]

[1]Department of Electrical and Computer Engineering, University of Massachusetts Lowell, Lowell, MA 01854, USA

[2]Department of Chemical Engineering, University of Massachusetts Lowell, Lowell, MA 01854, USA

[a)] Corresponding author Email: anhar_bhuiyan@uml.edu

**Abstract**

In this work, Ge-doped β-$Ga_2O_3$ homoepitaxial films were grown on native (010) β-$Ga_2O_3$ substrates using low-pressure chemical vapor deposition (LPCVD). Controlled n-type doping was achieved with room-temperature carrier concentrations ranging from $7.4 \times 10^{17}$ to $2.57 \times 10^{18}$ $\mathrm{cm}^{-3}$ and corresponding electron mobilities of 105-62 $cm^2/V \cdot s$. The films exhibited smooth surface morphology with RMS roughness values of 2.94-3.97 nm, while X-ray diffraction, Raman spectroscopy, and X-ray photoelectron spectroscopy confirmed phase-pure β-$Ga_2O_3$ with excellent crystalline quality and near-stoichiometric composition. Temperature-dependent Hall measurements on the film with a room-temperature carrier concentration of $7.4 \times 10^{17}$ $\mathrm{cm}^{-3}$ and mobility of 105 $cm^2/V \cdot s$ yielded a peak electron mobility of 234 $cm^2/V \cdot s$ at 116 K, while charge-neutrality and transport modeling revealed a dominant shallow donor level with an activation energy of 14 meV, confirming efficient electrical activation of Ge donors. Vertical Ni/β-$Ga_2O_3$ Schottky barrier diodes fabricated using the Ge-doped drift layer exhibited good rectifying behavior with a turn-on voltage of 0.74 V, an ideality factor of 1.32, a Schottky barrier height of 1.02 eV, and a specific on-resistance of 2.49 $m\Omega \cdot cm^2$. Capacitance-voltage measurements yielded a net donor concentration of $7.7 \times 10^{17}$ $\mathrm{cm}^{-3}$ and a Schottky barrier height of 1.13 eV, in good agreement with Hall and current-voltage measurements. These results demonstrate that LPCVD enables controllable Ge doping while maintaining high structural and electronic quality, establishing LPCVD-grown Ge-doped β-$Ga_2O_3$ as a promising platform for future high-voltage power electronic devices.

## I. Introduction

β-$Ga_2O_3$ has emerged as one of the most promising ultrawide bandgap semiconductors for next-generation high-power and high-voltage electronic applications owing to its large bandgap (4.8-4.9 eV), high theoretical critical breakdown field (6-8 MV/cm), and the availability of large-area melt-grown native substrates with excellent crystalline quality[1, 2]. Rapid advances in both epitaxial growth and device development have established β-$Ga_2O_3$ as a strong candidate for power conversion systems operating under extreme voltage, temperature, and radiation environments [2-4]. High-quality β-$Ga_2O_3$ epitaxial layers have been demonstrated using metal-organic chemical vapor deposition (MOCVD), halide vapor phase epitaxy (HVPE), molecular beam epitaxy (MBE), mist chemical vapor deposition (Mist-CVD), and low-pressure chemical vapor deposition (LPCVD) [5-28]. These growth approaches have enabled substantial progress in crystal quality, growth rate, doping control, and wafer-scale epitaxy. In parallel, β-$Ga_2O_3$ device technologies have advanced significantly, including the demonstration of high-voltage Schottky barrier diodes, p-n heterojunction diodes, and field-effect transistors with breakdown voltages ranging from several kilovolts to beyond 10 kV [29-44].

The performance of β-$Ga_2O_3$ power devices critically depends on precise control of n-type doping over a wide concentration range. Lightly doped drift layers are required to sustain high blocking voltages, whereas heavily doped regions are necessary to minimize contact resistance and access resistance. Among the available donor impurities, Si, Sn, and Ge have been most extensively investigated. Previous theoretical and experimental studies have shown that Si and Ge preferentially occupy tetrahedrally coordinated Ga(I) lattice sites, producing shallow donor levels and efficient room-temperature carrier activation, whereas Sn preferentially occupies octahedral Ga(II) sites and is often associated with deeper donor levels [45, 46]. First-principles calculations suggest that Ge exhibits a strong energetic preference for shallow donor configurations, potentially enabling higher free-electron concentrations than Si at comparable dopant concentrations. Furthermore, the close ionic size match between $Ge^{4+}$ and $Ga^{3+}$ facilitates substitutional incorporation into the β-$Ga_2O_3$ lattice with minimal structural distortion [47]. These characteristics make Ge a promising dopant for achieving both high conductivity and excellent crystalline quality in β-$Ga_2O_3$ epitaxial layers. To realize such controllably doped epitaxial layers, the choice of growth technique is also equally important. Among the available approaches, LPCVD has emerged as a promising method for β-$Ga_2O_3$ epitaxy because it employs elemental gallium and oxygen precursors rather than metal-organic precursors or chloride-based chemistries. This simplified growth environment can reduce precursor-related impurities while maintaining excellent control over growth conditions and doping incorporation. Previous studies have demonstrated high-quality LPCVD-grown β-$Ga_2O_3$ films with high growth rates reaching tens of μm/hr on both native β-$Ga_2O_3$ and sapphire substrates, together with successful incorporation of Si, and Sn, and Ge donors [16, 28, 48-53]. To date, Ge doping of β-$Ga_2O_3$ has been demonstrated using a variety of epitaxial growth techniques, including MBE, LPCVD and MOCVD homoepitaxy on native β-$Ga_2O_3$ substrates [45, [54, 55]. These studies reported electron mobilities ranging from approximately 90 to 140 cm²/V·s at carrier concentrations spanning the low-$10^{16}$ to low-$10^{18}$ $cm^{-3}$ range. In MBE-grown films [45], mobilities as high as 97 cm²/V·s were achieved at carrier concentrations near $1.6 \times 10^{18}$ $cm^{-3}$, while heavily doped layers approaching $10^{20}$ $cm^{-3}$ exhibited mobilities of approximately 38-40 cm²/V·s. Similarly, MOCVD-grown Ge-doped β-$Ga_2O_3$ demonstrated carrier concentrations from $2 \times 10^{16}$ to $3 \times 10^{20}$ $cm^{-3}$ with mobilities ranging from about 140 to 38 cm²/V·s [55]. LPCVD-grown Ge-doped β-$Ga_2O_3$ homoepitaxial films have also been reported, exhibiting carrier concentrations between $1.9 \times 10^{17}$ and $3 \times 10^{18}$ $cm^{-3}$ and room-temperature Hall mobilities of 20-43 cm²/V·s [55]. These studies confirm the effectiveness of Ge as a shallow donor in β-$Ga_2O_3$ and demonstrate its capability to provide controllable n-type conductivity over a wide carrier concentration range. However, comprehensive investigations correlating Ge incorporation, structural quality, surface morphology, chemical composition, electrical transport, and Schottky diode performance in LPCVD-grown Ge-doped β-$Ga_2O_3$ homoepitaxial films remain limited.

In this work, we report the homoepitaxial growth of Ge-doped β-$Ga_2O_3$ films on (010) β-$Ga_2O_3$ substrates using LPCVD. The structural, morphological, chemical, and electrical properties of the films were systematically investigated using X-ray diffraction, Raman spectroscopy, scanning electron microscopy, atomic force microscopy, X-ray photoelectron spectroscopy, Hall-effect measurements, and temperature-dependent transport analysis. The influence of Ge incorporation on crystal quality, carrier concentration, mobility, and donor activation behavior was studied in detail. To further evaluate the electronic quality of the epitaxial layers, vertical Schottky barrier diodes were fabricated and characterized through current-voltage and capacitance-voltage measurements. The results demonstrate efficient Ge donor activation, excellent crystalline quality, smooth surface morphology, and high-performance rectifying characteristics, establishing LPCVD as a promising approach for Ge doping of β-$Ga_2O_3$ for future high-voltage power electronic applications.

## II. Experimental Details

Ge-doped β-$Ga_2O_3$ epitaxial films were grown homoepitaxially on both Fe-doped (semi-insulating) and Sn-doped (conductive) (010) β-$Ga_2O_3$ substrates using a low-pressure chemical vapor deposition (LPCVD) system. The LPCVD reactor consists of a horizontal quartz tube housed within a three-zone furnace, with all growth zones maintained at 1100 °C during deposition. High-purity metallic gallium (99.99999%) and oxygen gas (99.999%) were employed as the gallium and oxygen precursors, respectively, while ultra-high-purity argon (99.9999%) was used as the carrier and purge gas. N-type doping was achieved using high-purity metallic germanium pellets (99.9999%) as the dopant source. Prior to growth, the β-$Ga_2O_3$ substrates were sequentially cleaned in acetone, isopropyl alcohol (IPA), and deionized (DI) water, followed by nitrogen drying. The Ga and Ge source materials were placed in separate quartz crucibles inside an inner quartz tube located within the growth zone. The Ge source was positioned upstream of the Ga source, while the substrate was placed downstream to facilitate efficient precursor transport and dopant incorporation during growth. The epitaxial films were deposited at a substrate temperature of 1100 °C under a chamber pressure of ~1.5 Torr with Ar and $O_2$ flow rates of 100 and 20 sccm, respectively. The Ga-Ge source spacing and Ga-substrate distance were maintained at 33 cm and 4.5 cm, respectively. The films exhibited room-temperature carrier concentrations ranging from $7.4 \times 10^{17}$ to $2.57 \times 10^{18}$ cm$^{-3}$ and Hall mobilities ranging from 105 to 62 cm$^2$/V·s. Film thicknesses between 1.05 and 2.54 μm were obtained by adjusting the growth duration between 20 and 60 min.

The surface morphology of the as-grown films was examined using field-emission scanning electron microscopy (FESEM, JEOL JSM-7401F) and atomic force microscopy (AFM, Park XE-100). Film thicknesses were determined from cross-sectional FESEM images on co-loaded sapphire substrates. The crystal structure, phase purity, and crystalline quality were evaluated using high-resolution X-ray diffraction (XRD) measurements performed on a Rigaku SmartLab diffractometer equipped with Cu Kα radiation (λ = 1.5418 Å). Raman spectroscopy was conducted using a Horiba LabRAM Evolution system with a 532 nm excitation laser to investigate lattice vibrational modes and confirm phase purity. The chemical composition and bonding states of the films were investigated using X-ray photoelectron spectroscopy (XPS) with a monochromatized Al Kα x-ray source (hν = 1486.6 eV). Electrical transport properties were characterized using the van der Pauw Hall configuration on an Ecopia HMS-5300 Hall measurement system over a temperature range of 80-350 K. Carrier concentration and electron mobility were extracted from room-temperature and temperature-dependent Hall measurements to evaluate Ge donor activation and carrier transport behavior. Vertical Schottky barrier diodes (SBDs) were fabricated on Ge-doped β-$Ga_2O_3$ epitaxial layers grown on Sn-doped (010) β-$Ga_2O_3$ substrates. Device fabrication began with the deposition of a Ti/Au (30/100 nm) metal stack on the backside of the substrate by electron-beam evaporation to form the Ohmic contact. The samples were subsequently annealed in a nitrogen atmosphere at 470 °C for 1 min using rapid thermal annealing. Schottky contacts were then defined on the epitaxial surface by photolithography, followed by deposition of a Ni/Au (20/100 nm) metal stack to form the Schottky anode. The electrical characteristics of the fabricated devices were evaluated using current density-voltage (J-V) and capacitance-voltage (C-V) measurements performed with a Keithley 4200A semiconductor parameter analyzer.

## III. Results and Discussion

The surface morphology of the LPCVD-grown Ge-doped β-$Ga_2O_3$ films was investigated using SEM and AFM, as shown in Figure 1. Figs. 1(a)-1(c) present the surface SEM images of films with thicknesses of 2.54, 1.05, and 2.54 μm, carrier concentrations of $7.4 \times 10^{17}$, $2.41 \times 10^{18}$, and $2.57 \times 10^{18}$ $cm^{-3}$, and corresponding room-temperature electron mobilities of 105, 87, and 62 $cm^2$/V·s, respectively. All samples exhibit smooth and continuous surfaces characterized by well-defined anisotropic surface striations oriented along the [001] crystallographic direction. To further evaluate the surface morphology, AFM measurements were performed over a $5 \times 5$ $\mu m^2$ scan area, as shown in Figs. 1(d)-1(f). All films exhibit elongated terrace-like surface features consistent with the SEM observations. The RMS roughness values were measured to be 2.94, 3.12, and 3.97 nm for carrier concentrations of $7.4 \times 10^{17}$, $2.41 \times 10^{18}$, and $2.57 \times 10^{18}$ $cm^{-3}$, respectively. A gradual increase in roughness is observed with increasing carrier concentration. Similar RMS roughness values have previously been reported for LPCVD-grown Si-doped β-$Ga_2O_3$ homoepitaxial films [28, 56], indicating that the incorporation of Ge does not significantly compromise the surface quality of the β-$Ga_2O_3$ films.

To evaluate the structural quality of the Ge-doped β-$Ga_2O_3$ homoepitaxial films, high-resolution XRD measurements were performed, as shown in Figure 2. Fig. 2(a) presents the wide-range ω-2θ scan of the film with a carrier concentration of $7.4 \times 10^{17}$ $cm^{-3}$. A single intense diffraction peak corresponding to the β-$Ga_2O_3$ (020) reflection is observed, confirming the successful homoepitaxial growth of (010)-oriented β-$Ga_2O_3$ on the native substrate. No additional diffraction peaks associated with other β-$Ga_2O_3$ orientations or metastable polymorphs are detected, indicating good phase purity and the absence of secondary phases. The crystalline quality of the Ge-doped films was further assessed using ω-rocking curve measurements of the β-$Ga_2O_3$

(020) reflection, as shown in Fig. 2(b). The corresponding full width at half maximum (FWHM) values were measured to be 97, 107, and 124 arcsec for carrier concentrations of $7.4 \times 10^{17}$, $2.41 \times 10^{18}$, and $2.57 \times 10^{18}$ $cm^{-3}$, respectively. A gradual increase in FWHM is observed with increasing carrier concentration. This behavior is likely associated with localized lattice distortions and increased structural disorder resulting from the substitutional incorporation of Ge atoms into Ga lattice sites. Nevertheless, only a modest increase in FWHM is observed despite a more than threefold increase in carrier concentration. The measured rocking-curve widths remain comparable to those previously reported for high-quality LPCVD-grown Si-doped $\beta$-$Ga_2O_3$ homoepitaxial films [28, 48, 57], demonstrating that LPCVD enables effective Ge donor incorporation while preserving good crystalline quality.

To investigate the chemical composition and bonding states of the Ge-doped $\beta$-$Ga_2O_3$ homoepitaxial films, XPS measurements were performed on the sample with a carrier concentration of $7.4 \times 10^{17}$ $cm^{-3}$, as shown in Figure 3. The survey spectrum in Fig. 3(a) reveals the characteristic Ga and O core-level features of $\beta$-$Ga_2O_3$. No additional peaks associated with metallic impurities, secondary phases, or other contaminant species are detected, indicating the high chemical purity of the LPCVD-grown $\beta$-$Ga_2O_3$ film. The high-resolution O 1s spectrum shown in Fig. 3(b) can be deconvoluted into two components centered near 530.8 and 532.2 eV. The dominant peak at around 530.8 eV is attributed to lattice oxygen bonded to Ga (Ga-O), characteristic of $\beta$-$Ga_2O_3$. The higher binding-energy component near 532.2 eV is associated with surface hydroxyl species (O-H) and adsorbed oxygen-containing surface groups. The Ga 3d spectrum shown in Fig. 3(c) is dominated by a peak centered near 20.3 eV, corresponding to Ga-O bonding within the $\beta$-$Ga_2O_3$ lattice. A weaker higher-binding-energy component is also observed and is attributed to surface hydroxyl (O-H) species. Quantitative analysis of the O 1s and Ga 3d spectra yielded atomic concentrations of 60.09% oxygen and 39.91% gallium, resulting in an O/Ga atomic ratio of 1.51. This value closely matches the stoichiometric ratio of 1.5 expected for β-$Ga_2O_3$, confirming that the LPCVD-grown Ge-doped film maintains near-ideal stoichiometry.

The vibrational properties and crystalline quality of the LPCVD-grown Ge-doped β-$Ga_2O_3$ homoepitaxial film were investigated using Raman spectroscopy, as shown in Figure 4. The primitive unit cell of monoclinic β-$Ga_2O_3$ contains 10 atoms, giving rise to 30 phonon modes, of which 27 are optical modes [58]. At the Γ-point, these optical modes belong to the irreducible representation $\Gamma^{opt} = 10Ag + 5Bg + 4Au + 8Bu$, where the Ag and Bg modes are Raman active, while the Au and Bu modes are infrared active. For measurements performed on the (010) surface, only Ag modes are expected to be observed because the Bg modes are symmetry forbidden under the employed scattering geometry. As shown in Fig. 4, the Raman spectrum exhibits all ten characteristic Ag phonon modes of β-$Ga_2O_3$, including Ag(1), Ag(2), Ag(3), Ag(4), Ag(5), Ag(6), Ag(7), Ag(8), Ag(9), and Ag(10), located at approximately 111, 170, 200, 320, 347, 416, 475, 630, 659, and 767 $cm^{-1}$, respectively. The measured peak positions are in excellent agreement with previously reported experimental and theoretical studies of β-$Ga_2O_3$, confirming the formation of phase-pure monoclinic β-$Ga_2O_3$ [28, 50, 57, 58]. The observed Raman modes represent different lattice vibrations within the β-$Ga_2O_3$ crystal structure. The low-frequency modes near 111, 170, and 200 $cm^{-1}$ are associated with translational and librational motions of the $GaO_4$ tetrahedra and $GaO_6$ octahedra network. The intermediate-frequency modes located between 320 and 475 $cm^{-1}$ originate primarily from deformations and bending vibrations of the $GaO_6$ octahedra. The high-frequency modes near 630, 659, and 767 $cm^{-1}$ are attributed to bending and stretching vibrations of the Ga-O bonds within the $GaO_4$ tetrahedral units. The presence of these well-defined vibrational modes indicates that the β-$Ga_2O_3$ crystal structure is preserved after Ge incorporation.

Furthermore, the absence of additional Raman modes confirms that Ge incorporation preserves the phase purity and structural integrity of the β-$Ga_2O_3$ lattice.

To further investigate the donor activation behavior and carrier transport mechanisms in the LPCVD-grown Ge-doped β-$Ga_2O_3$ homoepitaxial films, temperature-dependent Hall-effect measurements were analyzed using a self-consistent charge-neutrality model combined with Boltzmann transport calculations [59-61]. The measured carrier concentration and Hall mobility, together with the corresponding model fits, are shown in Fig. 5. Details of the complete transport model and scattering formalism are provided in the Supplementary Material. Figure 5(a) shows the temperature dependence of the carrier concentration and Hall mobility for the Ge-doped β-$Ga_2O_3$ homoepitaxial film grown with $7.4 \times 10^{17}\ \mathrm{cm}^{-3}$ carrier concentration. At room temperature, the film exhibits a Hall mobility of $105\ \mathrm{cm}^2/\mathrm{V}\cdot\mathrm{s}$. As the temperature decreases, the carrier concentration gradually decreases to $2.06 \times 10^{17}\ \mathrm{cm}^{-3}$ at 80 K, indicating donor freeze-out. Simultaneously, the Hall mobility increases and reaches a peak value of $234\ \mathrm{cm}^2/\mathrm{V}\cdot\mathrm{s}$ at 116 K due to the reduction of phonon scattering at lower temperatures. The temperature dependence of the electron concentration was analyzed using the charge-neutrality condition [59-62]

$$n(T) + N_A + \frac{N_{\mathrm{line}}}{d} \quad = \quad N_{D1}^{+}(T) + N_{D2}^{+}(T), \qquad (1)$$

where n(T) is the free electron concentration, $N_A$ is the compensating acceptor concentration, $N_{\mathrm{line}}$ represents the effective density of charged line defects, *d* is the spacing between charged centers along the defect line, and $N_{D1}^{+}$and $N_{D2}^{+}$ are the ionized donor concentrations associated with two donor states. The donor ionization was described using Fermi-Dirac statistics and solved self-consistently with the charge-neutrality equation. As shown in Fig. 5(b), excellent agreement is obtained between the experimental Hall data and the calculated carrier concentration. The fitting results reveal two donor populations with concentrations of $1.20 \times 10^{18}$ and $5 \times 10^{16}\ \mathrm{cm}^{-3}$ and corresponding activation energies of 14 and 100 meV, respectively, as summarized in Table 1. The shallow donor level dominates the room-temperature transport and is consistent with the expected behavior of substitutional Ge donors in β-$Ga_2O_3$ [59]. The second donor state, although present at a significantly lower concentration, contributes to the temperature dependence of the carrier concentration The extracted acceptor concentration remains low ($N_A = 1 \times 10^{15}$ cm$^{-3}$). The Hall mobility was analyzed using Matthiessen's rule [59, 61-63],

$$\frac{1}{\mu_{\text{tot}}} = \frac{1}{\mu_{\text{POP}}} + \frac{1}{\mu_{\text{II}}} + \frac{1}{\mu_{\text{NI}}} + \frac{1}{\mu_{\text{ADP}}} + \frac{1}{\mu_{\text{line}}}, \quad (2)$$

where $\mu_{\text{POP}}$, $\mu_{\text{II}}$, $\mu_{\text{NI}}$, $\mu_{\text{ADP}}$, and $\mu_{\text{line}}$ represent the mobility contributions limited by polar optical phonon scattering, ionized impurity scattering, neutral impurity scattering, acoustic deformation potential scattering, and charged line-defect scattering, respectively. The fitting yielded an effective charged line-defect density of $5.05 \times 10^8\ cm^{-2}$. This parameter represents the density of extended charged scattering centers incorporated in the transport model to account for the residual low-temperature mobility degradation. The calculated mobility behavior indicates that polar optical phonon scattering dominates at elevated temperatures, whereas impurity- and defect-related scattering mechanisms become increasingly important at lower temperatures. The excellent agreement between the measured and calculated carrier concentration and mobility demonstrates that the adopted model successfully captures the dominant transport processes in the Ge-doped β-$Ga_2O_3$ homoepitaxial films.

To evaluate the device-level quality of the LPCVD-grown Ge-doped β-$Ga_2O_3$ homoepitaxial layer, vertical Schottky barrier diodes were fabricated on a 2.54 μm thick $7.7 \times 10^{17}$ cm$^{-3}$ Ge-doped β-$Ga_2O_3$ drift layer grown on a conductive Sn-doped (010) β-$Ga_2O_3$ substrate. Figure 6(a) shows the schematic cross-section of the fabricated diode structure, consisting of a Ni/Au Schottky anode on the Ge-doped β-$Ga_2O_3$ drift layer and a Ti/Au backside Ohmic contact on the Sn-doped $n^+$ β-

$Ga_2O_3$ substrate. Figure 6(b) shows the room-temperature current density-voltage (J-V) characteristics of the fabricated SBD. The device exhibits clear rectifying behavior with a turn-on voltage of 0.74 V and a low specific on-resistance of $2.49\ \mathrm{m\Omega \cdot cm^2}$. The forward-bias characteristics were analyzed using the thermionic emission model [64, 65],

$$J = J_s\left[exp\left(\frac{qV}{\eta k_0 T}\right) - 1\right] \qquad (3)$$

where $J_s$ is the reverse saturation current density, $k_0$ is the Boltzmann constant, $q$ is the elementary charge, and $\eta$ is the ideality factor. The reverse saturation current density, $J_s$, is related to the Schottky barrier height through the Richardson thermionic-emission relation

$$J_s = A^* T^2\, exp\left(-\frac{q\varphi_B^{JV}}{k_0 T}\right) \qquad (4)$$

where $\varphi_B^{JV}$ is the apparent Schottky barrier height extracted from the J-V characteristics and $A^*$ is the effective Richardson constant ($A^* = \frac{4\pi q m_n^* k_0^2}{h^3}$) which is calculated to be 41.04 A $cm^{-2}$ $K^{-2}$ [64, 65]. The extracted ideality factor obtained from the J-V fitting is 1.32. Since the barrier height extracted from Eq. (4) depends sensitively on deviations of the ideality factor from unity, a correction was applied following the method proposed by Wagner et al. to obtain a more physically representative Schottky barrier height [66],

$$\varphi_B = \left(\varphi_B^{JV} - \frac{\eta - 1}{\eta}\frac{kT}{q}\, ln\frac{N_C}{N_D}\right)\eta \qquad (5)$$

where $N_C$ is the effective density of states in the conduction band and $N_D$ is the donor concentration in the drift layer. the corrected Schottky barrier height was determined to be 1.02 eV. Figure 6(c) shows the room-temperature capacitance-voltage (C-V) and corresponding $1/C^2 - V$ characteristics of the Ge-doped β-$Ga_2O_3$ Schottky barrier diode. The net donor concentration ($N_D^+ - N_A^-$), built-in potential $V_{bi}$, and Schottky barrier height $\Phi_B$ were extracted using the standard relations for Schottky diodes [64],

$$N_d^+ - N_a^- = \frac{2}{q\varepsilon_r\varepsilon_0 A^2\left(\frac{d\frac{1}{C^2}}{dV}\right)} \quad (6)$$

$$\frac{A^2}{C^2} = \left[\frac{2}{q\varepsilon_r\varepsilon_0(N_d^+ - N_a^-)}\right]\left(V_{bi} - \frac{kT}{q} - V\right) \quad (7)$$

$$\varphi_B = V_{bi} + \frac{KT}{q} ln\left[\frac{N_C}{N_d^+ - N_a^-}\right] \quad (8)$$

where $\varepsilon_r$=10 is the relative permittivity of β-$Ga_2O_3$, $\varepsilon_0$ is the vacuum permittivity, $A$ is the Schottky contact area, and $N_C$=5.2 × $10^{18}$ cm$^{-3}$ is the effective density of states in the conduction band. From the linear fit to the $1/C^2 - V$ characteristics, $V_{bi} - k_BT/q$ was extracted to be 0.98 V, corresponding to a built-in potential of $V_{bi}$=1.01 V at room temperature. Using Eq. (8), the Schottky barrier height determined from the C-V analysis was found to be 1.13 eV. This value is slightly higher than the barrier height extracted from the J-V characteristics (1.02 eV), which is commonly observed in β-$Ga_2O_3$ Schottky diodes and is generally attributed to barrier inhomogeneity, interface states, and the different transport mechanisms probed by J-V and C-V measurements [67]. The depth-dependent carrier concentration profile extracted from the C-V data is shown in Fig. 6(d). The average net donor concentration was determined to be 7.7 × $10^{17}$ cm$^{-3}$, which is in excellent agreement with the room-temperature Hall carrier concentration of 7.4 × $10^{17}$ cm$^{-3}$. Furthermore, the carrier concentration remains uniform over the depletion region.

## IV. Conclusion

In summary, Ge-doped β-$Ga_2O_3$ homoepitaxial films were grown on (010) β-$Ga_2O_3$ substrates using LPCVD and systematically investigated through structural, chemical, electrical, and device characterization. The films exhibited smooth surface morphology, narrow XRD rocking-curve widths, characteristic β-$Ga_2O_3$ Raman modes, and near-stoichiometric O/Ga composition, confirming the high structural and chemical quality of the LPCVD-grown Ge-doped $\beta$-$Ga_2O_3$ homoepitaxial layers. Temperature-dependent Hall measurements and transport modeling confirmed that Ge behaves predominantly as a shallow donor with a low activation energy, enabling efficient room-temperature carrier activation while maintaining promising carrier transport properties. The successful demonstration of vertical Schottky barrier diodes further validates the electronic quality of the LPCVD-grown Ge-doped $\beta$-$Ga_2O_3$ layers. The close agreement between Hall- and C-V-extracted carrier concentrations, together with the near-ideal diode characteristics, indicates that the incorporated Ge donors are electrically active and uniformly distributed within the epitaxial layer. Future optimization of LPCVD growth conditions may enable extension of the achievable doping range toward lower carrier concentrations required for high-voltage drift layers. Overall, these results establish LPCVD as a viable growth platform for Ge doping of $\beta$-$Ga_2O_3$ and highlight its potential for the development of future high-voltage $\beta$-$Ga_2O_3$ power devices requiring precise doping control, high crystalline quality, and device-grade epitaxial layers.

**Acknowledgement**

The authors acknowledge the funding support from the National Science Foundation (NSF) under award numbers ECCS-2532898 and ECCS-2501623 and Nuclear Regulatory Commission (NRC) under Grant No. 31310024M0046.

**Data Availability**

The data that support the findings of this study are available from the corresponding author upon reasonable request.

**Conflict of Interest**

The authors have no conflicts to disclose.

## References

(1) Higashiwaki, M.; Sasaki, K.; Kuramata, A.; Masui, T.; Yamakoshi, S. Gallium oxide ($Ga_2O_3$) metal-semiconductor field-effect transistors on single-crystal β-$Ga_2O_3$ (010) substrates. *Appl. Phys. Lett.* **2012**, *100* (1), 013504. DOI: 10.1063/1.3674287.

(2) Higashiwaki, M.; Wong, M. H. Beta-gallium oxide material and device technologies. *Annual Review of Materials Research* **2024**, *54* (1), 175-198.

(3) Pearton, S.; Aitkaliyeva, A.; Xian, M.; Ren, F.; Khachatrian, A.; Ildefonso, A.; Islam, Z.; Rasel, M. A. J.; Haque, A.; Polyakov, A. Y. Radiation damage in wide and ultra-wide bandgap semiconductors. *ECS Journal of Solid State Science and Technology* **2021**, *10* (5), 055008.

(4) Islam, A. E.; Sepelak, N. P.; Liddy, K. J.; Kahler, R.; Williams, J.; Dryden, D. M.; Green, A. J.; Chabak, K. D. High temperature operation of beta-Ga 2 O 3 transistors. *IMAPSource Proc.* **2023**, *2022*, 000059-000062.

(5) Feng, Z.; Bhuiyan, A.; Karim, M. R.; Zhao, H. MOCVD homoepitaxy of Si-doped (010) β-$Ga_2O_3$ thin films with superior transport properties. *Appl. Phys. Lett.* **2019**, *114* (25), 250601. DOI: 10.1063/1.5109678.

(6) Feng, Z.; Bhuiyan, A.; Xia, Z.; Moore, W.; Chen, Z.; McGlone, J. F.; Daughton, D. R.; Arehart, A. R.; Ringel, S. A.; Rajan, S.; et al. Probing Charge Transport and Background Doping in Metal-Organic Chemical Vapor Deposition-Grown (010) β-$Ga_2O_3$. *Phys. Status Solidi RRL* **2020**, *14* (8), 2000145. DOI: 10.1002/pssr.202000145.

(7) Peterson, C.; Bhattacharyya, A.; Chanchaiworawit, K.; Kahler, R.; Roy, S.; Liu, Y.; Rebollo, S.; Kallistova, A.; Mates, T. E.; Krishnamoorthy, S. 200 $cm^2$/Vs electron mobility and controlled low $10^{15}$ $cm^{-3}$ Si doping in (010) β-$Ga_2O_3$ epitaxial drift layers. *Appl. Phys. Lett.* **2024**, *125* (18), 182103. DOI: 10.1063/5.0230413.

(8) Bhuiyan, A.; Feng, Z.; Johnson, J. M.; Chen, Z.; Huang, H.-L.; Hwang, J.; Zhao, H. MOCVD epitaxy of β-$(Al_xGa_{1-x})_2O_3$ thin films on (010) $Ga_2O_3$ substrates and N-type doping. *Appl. Phys. Lett.* **2019**, *115* (12), 120602. DOI: 10.1063/1.5123495.

(9) Bhuiyan, A.; Meng, L.; Huang, H.-L.; Sarker, J.; Chae, C.; Mazumder, B.; Hwang, J.; Zhao, H. Metalorganic chemical vapor deposition of β-$(Al_xGa_{1-x})_2O_3$ thin films on (001) β-$Ga_2O_3$ substrates. *APL Mater.* **2023**, *11* (4), 041112. DOI: 10.1063/5.0142746.

(10) Bhuiyan, A.; Feng, Z.; Meng, L.; Zhao, H. Tutorial: Metalorganic chemical vapor deposition of β-$Ga_2O_3$ thin films, alloys, and heterostructures. *J. Appl. Phys.* **2023**, *133* (21), 211103. DOI: 10.1063/5.0147787.

(11) Bhuiyan, A.; Feng, Z.; Huang, H.-L.; Meng, L.; Hwang, J.; Zhao, H. Metalorganic chemical vapor deposition of α-$Ga_2O_3$ and α-$(Al_xGa_{1-x})_2O_3$ thin films on m-plane sapphire substrates. *APL Mater.* **2021**, *9* (10), 101109. DOI: 10.1063/5.0065087.

(12) Zheng, Y.; Feng, Z.; Bhuiyan, A.; Meng, L.; Dhole, S.; Jia, Q.; Zhao, H.; Seo, J.-H. Large-size free-standing single-crystal β-Ga 2 O 3 membranes fabricated by hydrogen implantation and lift-off. *Journal of Materials Chemistry C* **2021**, *9* (19), 6180-6186.

(13) Saha, S.; Meng, L.; Bhuiyan, A.; Sharma, A.; Saha, C. N.; Zhao, H.; Singisetti, U. Electrical characteristics of in situ Mg-doped β-$Ga_2O_3$ current-blocking layer for vertical devices. *Appl. Phys. Lett.* **2023**, *123* (13), 132105. DOI: 10.1063/5.0155882.

(14) Sarker, S.; Khan, S. A.; Ibreljic, A.; Bhuiyan, A. HVPE Growth of Si-Doped β-$Ga_2O_3$ on Sapphire: Influence of Substrate Offcut on Structural and Electrical Properties. *arXiv preprint arXiv:2606.05607* **2026**.

(15) Wang, G.; Xie, S.; Brand, W.; Khan, S. A.; Ibreljic, A.; Shima, D.; Ma, Y.; Challa, B.;

Sanyal, S.; Alema, F.; et al. Single-Crystalline High-Quality β-$Ga_2O_3$ Pseudosubstrate on Sapphire through Sputtering for Epitaxial Deposition. *ACS Applied Engineering Materials* **2026**. DOI: 10.1021/acsaenm.6c00439.

(16) Khan, S. A.; Ibreljic, A.; Bhuiyan, A. Vertical β-$Ga_2O_3$ Schottky diodes with LPCVD-grown Sn-doped drift layer. *APL Electron. Devices* **2026**, *2* (2), 026112. DOI: 10.1063/5.0319027.

(17) Bhuiyan, A.; Meng, L.; Yu, D. S.; Dhara, S.; Huang, H.-L.; Vangipuram, V. G. T.; Hwang, J.; Rajan, S.; Zhao, H. Electrical and structural characterization of in situ MOCVD $Al_2O_3$/β-$Ga_2O_3$ and $Al_2O_3$/β-$(Al_xGa_{1-x})_2O_3$ MOSCAPs. *J. Appl. Phys.* **2025**, *137* (17), 174101. DOI: 10.1063/5.0256525.

(18) Saha, S.; Meng, L.; Yu, D. S.; Bhuiyan, A.; Zhao, H.; Singisetti, U. High growth rate metal organic chemical vapor deposition grown $Ga_2O_3$ (010) Schottky diodes. *J. Vac. Sci. Technol. A* **2024**, *42* (4), 042705. DOI: 10.1116/6.0003533.

(19) McGlone, J. F.; Ghadi, H.; Cornuelle, E.; Armstrong, A.; Burns, G.; Feng, Z.; Bhuiyan, A.; Zhao, H.; Arehart, A. R.; Ringel, S. A. Proton radiation effects on electronic defect states in MOCVD-grown (010) β-$Ga_2O_3$. *J. Appl. Phys.* **2023**, *133* (4), 045702. DOI: 10.1063/5.0121416.

(20) Bhuiyan, A.; Feng, Z.; Meng, L.; Fiedler, A.; Huang, H.-L.; Neal, A. T.; Steinbrunner, E.; Mou, S.; Hwang, J.; Rajan, S.; et al. Si doping in MOCVD grown (010) β-$(Al_xGa_{1-x})_2O_3$ thin films. *J. Appl. Phys.* **2022**, *131* (14), 145301. DOI: 10.1063/5.0084062.

(21) Meng, L.; Feng, Z.; Bhuiyan, A.; Zhao, H. High-Mobility MOCVD β-$Ga_2O_3$ Epitaxy with Fast Growth Rate Using Trimethylgallium. *Cryst. Growth Des.* **2022**, *22* (6), 3896-3904. DOI: 10.1021/acs.cgd.2c00290.

(22) Bhuiyan, A.; Feng, Z.; Huang, H.-L.; Meng, L.; Hwang, J.; Zhao, H. MOCVD growth and band offsets of κ-phase $Ga_2O_3$ on c-plane sapphire, GaN- and AlN-on-sapphire, and (100) YSZ substrates. *J. Vac. Sci. Technol. A* **2022**, *40* (6), 062704. DOI: 10.1116/6.0002106.

(23) Bhuiyan, A.; Feng, Z.; Meng, L.; Zhao, H. MOCVD growth of (010) β-$(Al_xGa_{1-x})_2O_3$ thin films. *Journal of Materials Research* **2021**, *36* (23), 4804-4815. DOI: 10.1557/s43578-021-00354-8.

(24) Bhuiyan, A.; Feng, Z.; Johnson, J. M.; Huang, H.-L.; Hwang, J.; Zhao, H. MOCVD growth of β-phase $(Al_xGa_{1-x})_2O_3$ on ( 2 ¯01) β-$Ga_2O_3$ substrates. *Appl. Phys. Lett.* **2020**, *117* (14), 142107. DOI: 10.1063/5.0025478.

(25) Feng, Z.; Bhuiyan, A.; Kalarickal, N. K.; Rajan, S.; Zhao, H. Mg acceptor doping in MOCVD (010) β-$Ga_2O_3$. *Appl. Phys. Lett.* **2020**, *117* (22), 222106. DOI: 10.1063/5.0031562.

(26) Bhuiyan, A.; Feng, Z.; Johnson, J. M.; Huang, H.-L.; Hwang, J.; Zhao, H. MOCVD Epitaxy of Ultrawide Bandgap β-$(Al_xGa_{1-x})_2O_3$ with High-Al Composition on (100) β-$Ga_2O_3$ Substrates. *Cryst. Growth Des.* **2020**, *20* (10), 6722-6730. DOI: 10.1021/acs.cgd.0c00864.

(27) Kaneko, K.; Uno, K.; Jinno, R.; Fujita, S. Prospects for phase engineering of semi-stable $Ga_2O_3$ semiconductor thin films using mist chemical vapor deposition. *J. Appl. Phys.* **2022**, *131* (9), 090902. DOI: 10.1063/5.0069554.

(28) Khan, S. A.; Ibreljic, A.; Margiotta, S.; Bhuiyan, A. Low-pressure CVD grown Si-doped β-$Ga_2O_3$ films with promising electron mobilities and high growth rates. *Appl. Phys. Lett.* **2025**, *126* (1), 012103. DOI: 10.1063/5.0245559.

(29) Wakimoto, D.; Lin, C.-H.; Ema, K.; Ueda, Y.; Miyamoto, H.; Sasaki, K.; Kuramata, A. A multi-fin normally-off β-$Ga_2O_3$ vertical transistor with a breakdown voltage exceeding 10 kV. *Appl. Phys. Express* **2025**, *18* (10), 106502.

(30) Li, W.; Hu, Z.; Nomoto, K.; Zhang, Z.; Hsu, J.-Y.; Thieu, Q. T.; Sasaki, K.; Kuramata, A.; Jena, D.; Xing, H. G. 1230 V β-$Ga_2O_3$ trench Schottky barrier diodes with an ultra-low leakage

current of <1 μA/cm2. *Appl. Phys. Lett.* **2018**, *113* (20), 202101. DOI: 10.1063/1.5052368.
(31) Li, J.-S.; Chiang, C.-C.; Wan, H.-H.; Ren, F.; Liao, Y.-T.; Pearton, S. J. kV-class Ga2O3 vertical rectifiers fabricated on 4-in. diameter substrates. *J. Vac. Sci. Technol. A* **2024**, *43* (1), 012701. DOI: 10.1116/6.0004141.
(32) Higashiwaki, M.; Sasaki, K.; Kamimura, T.; Hoi Wong, M.; Krishnamurthy, D.; Kuramata, A.; Masui, T.; Yamakoshi, S. Depletion-mode Ga2O3 metal-oxide-semiconductor field-effect transistors on β-Ga2O3 (010) substrates and temperature dependence of their device characteristics. *Appl. Phys. Lett.* **2013**, *103* (12), 123511.
(33) Wong, M. H.; Sasaki, K.; Kuramata, A.; Yamakoshi, S.; Higashiwaki, M. Field-plated Ga 2 O 3 MOSFETs with a breakdown voltage of over 750 V. *IEEE Electron Device Lett.* **2015**, *37* (2), 212-215.
(34) Wong, M. H.; Takeyama, A.; Makino, T.; Ohshima, T.; Sasaki, K.; Kuramata, A.; Yamakoshi, S.; Higashiwaki, M. Radiation hardness of β-Ga2O3 metal-oxide-semiconductor field-effect transistors against gamma-ray irradiation. *Appl. Phys. Lett.* **2018**, *112* (2), 023503. DOI: 10.1063/1.5017810.
(35) Dhara, S.; Dheenan, A.; Rajan, S. Charge recovery by vacuum annealing in β-Ga2O3 multi-fin trench Schottky barrier diodes. *APL Electron. Devices* **2025**, *1* (2), 026122. DOI: 10.1063/5.0270286.
(36) Bhattacharyya, A.; Ranga, P.; Roy, S.; Peterson, C.; Alema, F.; Seryogin, G.; Osinsky, A.; Krishnamoorthy, S. Multi-kV Class β-$Ga_2O_3$ MESFETs With a Lateral Figure of Merit Up to 355 MW/cm$^2$. *IEEE Electron Device Letters* **2021**, *42* (9), 1272-1275.
(37) Ren, Z.; Huang, H.-C.; Lee, H.; Chan, C.; Roberts, H. C.; Wu, X.; Waseem, A.; Bhuiyan, A.; Zhao, H.; Zhu, W.; et al. Temperature dependent characteristics of β-Ga2O3 FinFETs by MacEtch. *Appl. Phys. Lett.* **2023**, *123* (4), 043505. DOI: 10.1063/5.0159420.
(38) Kalarickal, N. K.; Feng, Z.; Bhuiyan, A.; Xia, Z.; Moore, W.; McGlone, J. F.; Arehart, A. R.; Ringel, S. A.; Zhao, H.; Rajan, S. Electrostatic engineering using extreme permittivity materials for ultra-wide bandgap semiconductor transistors. *IEEE Trans. Electron Devices* **2020**, *68* (1), 29-35.
(39) Huang, H.-C.; Ren, Z.; Bhuiyan, A.; Feng, Z.; Yang, Z.; Luo, X.; Huang, A. Q.; Green, A.; Chabak, K.; Zhao, H. β-Ga2O3 FinFETs with ultra-low hysteresis by plasma-free metal-assisted chemical etching. *Appl. Phys. Lett.* **2022**, *121* (5), 052102.
(40) Saha, C. N.; Roy, S.; Liu, Y.; Peterson, C.; Krishnamoorthy, S. 2.34 kV β-Ga2O3 vertical trench RESURF Schottky barrier diode with sub-micron fin width. *APL Electron. Devices* **2025**, *1* (4), 046125. DOI: 10.1063/5.0299732.
(41) Peterson, C.; Saha, C. N.; Kahler, R.; Liu, Y.; Mattapalli, A.; Roy, S.; Krishnamoorthy, S. Kilovolt-class β-Ga2O3 field-plated Schottky barrier diodes with MOCVD-grown intentionally 1015 cm−3 doped drift layers. *J. Appl. Phys.* **2025**, *138* (18), 185105. DOI: 10.1063/5.0302735.
(42) Gilankar, A.; Katta, A.; Das, N.; Kalarickal, N. K. >3kV NiO/Ga2O3 Heterojunction Diodes With Space-Modulated Junction Termination Extension and Sub-1V Turn-On. *IEEE Journal of the Electron Devices Society* **2025**, *13*, 373-377. DOI: 10.1109/JEDS.2025.3562028.
(43) Liu, Y.; Peterson, C.; Saha, C. N.; Tadjer, M. J.; Krishnamoorthy, S. VBr> 10 kV E-Beam/Sputtered Vertical NiOx/(011)\beta-Ga2O3 HJDs with PFOM> 2.3 GW/cm2. *arXiv preprint arXiv:2604.27262* **2026**.
(44) Khan, S. A.; Saha, S.; Singisetti, U.; Bhuiyan, A. Radiation resilience of β-Ga2O3 Schottky barrier diodes under high dose gamma radiation. *J. Appl. Phys.* **2024**, *136* (22), 225701. DOI: 10.1063/5.0233995.

(45) Alema, F.; Seryogin, G.; Osinsky, A.; Osinsky, A. Ge doping of β-Ga2O3 by MOCVD. *APL Mater.* **2021**, *9* (9), 091102. DOI: 10.1063/5.0059657.
(46) *J. S. Speck, in AVS 66th International Symposium and Exhibition, Columbus, OH, 2019.*
(47) Víllora, E. G.; Shimamura, K.; Yoshikawa, Y.; Ujiie, T.; Aoki, K. Electrical conductivity and carrier concentration control in β-Ga2O3 by Si doping. *Appl. Phys. Lett.* **2008**, *92* (20), 202120. DOI: 10.1063/1.2919728.
(48) Rafique, S.; Karim, M. R.; Johnson, J. M.; Hwang, J.; Zhao, H. LPCVD homoepitaxy of Si doped β-Ga2O3 thin films on (010) and (001) substrates. *Appl. Phys. Lett.* **2018**, *112* (5), 052104. DOI: 10.1063/1.5017616.
(49) Joishi, C.; Rafique, S.; Xia, Z.; Han, L.; Krishnamoorthy, S.; Zhang, Y.; Lodha, S.; Zhao, H.; Rajan, S. Low-pressure CVD-grown β-Ga2O3bevel-field-plated Schottky barrier diodes. *Appl. Phys. Express* **2018**, *11* (3), 031101. DOI: 10.7567/apex.11.031101.
(50) Rafique, S.; Han, L.; Neal, A. T.; Mou, S.; Boeckl, J.; Zhao, H. Towards High-Mobility Heteroepitaxial β-Ga2O3 on Sapphire− Dependence on The Substrate Off-Axis Angle. *physica status solidi (a)* **2018**, *215* (2), 1700467.
(51) Rafique, S.; Han, L.; Zhao, H. Synthesis of wide bandgap Ga2O3 (Eg ~ 4.6–4.7 eV) thin films on sapphire by low pressure chemical vapor deposition. *Phys. Status Solidi A* **2016**, *213* (4), 1002-1009. DOI: https://doi.org/10.1002/pssa.201532711.
(52) Khan, S. A.; Ibreljic, A.; Bhuiyan, A. Plasma damage-free in situ etching of β-Ga2O3 using solid-source gallium in the LPCVD system. *Appl. Phys. Lett.* **2025**, *127* (10), 102105. DOI: 10.1063/5.0277909.
(53) Khan, S. A.; Ibreljic, A.; Bhuiyan, A. Quasi-vertical β-Ga2O3 Schottky diodes on sapphire using all-LPCVD growth and plasma-free Ga-assisted etching. *APL Electron. Devices* **2025**, *1* (3), 036125. DOI: 10.1063/5.0280191.
(54) Ahmadi, E.; Koksaldi, O. S.; Kaun, S. W.; Oshima, Y.; Short, D. B.; Mishra, U. K.; Speck, J. S. Ge doping of β-Ga2O3 films grown by plasma-assisted molecular beam epitaxy. *Appl. Phys. Express* **2017**, *10* (4), 041102. DOI: 10.7567/APEX.10.041102.
(55) Ranga, P.; Bhattacharyya, A.; Whittaker-Brooks, L.; Scarpulla, M. A.; Krishnamoorthy, S. N-type doping of low-pressure chemical vapor deposition grown β-Ga2O3 thin films using solid-source germanium. *J. Vac. Sci. Technol. A* **2021**, *39* (3), 030404. DOI: 10.1116/6.0001004.
(56) Zhang, Y.; Feng, Z.; Karim, M. R.; Zhao, H. High-temperature low-pressure chemical vapor deposition of β-Ga2O3. *J. Vac. Sci. Technol. A* **2020**, *38* (5), 050806. DOI: 10.1116/6.0000360.
(57) Rafique, S.; Han, L.; Tadjer, M. J.; Freitas, J. A.; Mahadik, N. A.; Zhao, H. Homoepitaxial growth of β-Ga2O3 thin films by low pressure chemical vapor deposition. *Appl. Phys. Lett.* **2016**, *108* (18), 182105. DOI: 10.1063/1.4948944.
(58) Kranert, C.; Sturm, C.; Schmidt-Grund, R.; Grundmann, M. Raman tensor elements of β-Ga2O3. *Scientific Reports* **2016**, *6* (1), 35964. DOI: 10.1038/srep35964.
(59) Neal, A. T.; Mou, S.; Rafique, S.; Zhao, H.; Ahmadi, E.; Speck, J. S.; Stevens, K. T.; Blevins, J. D.; Thomson, D. B.; Moser, N.; et al. Donors and deep acceptors in β-Ga2O3. *Appl. Phys. Lett.* **2018**, *113* (6), 062101. DOI: 10.1063/1.5034474.
(60) Lundstrom, M. S. Fundamentals of Carrier Transport, 2nd edn. *Measurement Science and Technology* **2002**, *13*, 230 - 230.
(61) Wakamatsu, T.; Izumi, H.; Tanake, H.; Isobe, Y.; Kaneko, K.; Tanaka, K. Mist-chemical vapor deposition grown Ge-doped α-Ga2O3 thin films with high electron mobility. *Appl. Phys. Lett.* **2026**, *128* (1), 012105. DOI: 10.1063/5.0308320.

(62) Takane, H.; Izumi, H.; Hojo, H.; Wakamatsu, T.; Tanaka, K.; Kaneko, K. Effect of dislocations and impurities on carrier transport in α-Ga2O3 on m-plane sapphire substrate. *Journal of Materials Research* **2023**, *38* (10), 2645-2654. DOI: 10.1557/s43578-023-01015-8.
(63) Ma, N.; Tanen, N.; Verma, A.; Guo, Z.; Luo, T.; Xing, H.; Jena, D. Intrinsic electron mobility limits in β-Ga2O3. *Appl. Phys. Lett.* **2016**, *109* (21), 212101.
(64) Sze, S. M.; Li, Y.; Ng, K. K. *Physics of Semiconductor Devices*; Wiley & sons, 2021.
(65) Cheung, S. K.; Cheung, N. W. Extraction of Schottky diode parameters from forward current-voltage characteristics. *Appl. Phys. Lett.* **1986**, *49* (2), 85-87. DOI: 10.1063/1.97359.
(66) Wagner, L.; Young, R.; Sugerman, A. A note on the correlation between the Schottky-diode barrier height and the ideality factor as determined from IV measurements. *IEEE Electron Device Lett.* **2005**, *4* (9), 320-322.
(67) Farzana, E.; Zhang, Z.; Paul, P. K.; Arehart, A. R.; Ringel, S. A. Influence of metal choice on (010) β-Ga2O3 Schottky diode properties. *Appl. Phys. Lett.* **2017**, *110* (20), 202102. DOI: 10.1063/1.4983610.

**Table 1**

Donor activation energies, donor concentrations, effective charged line-defect density, and compensating acceptor concentration extracted from self-consistent charge-neutrality and transport modeling of the temperature-dependent Hall-effect measurements for the Ge-doped β-$Ga_2O_3$ homoepitaxial film.

| $N_{D1}$ (cm⁻³) | $E_{D1}$ (meV) | $N_{D2}$ (cm⁻³) | $E_{D2}$ (meV) | $N_{line}$ (cm⁻²) | $N_A$ (cm⁻³) |
|---|---|---|---|---|---|
| $1.20 \times 10^{18}$ | 14 | $5 \times 10^{16}$ | 100 | $5.05 \times 10^{8}$ | $1 \times 10^{15}$ |

**Figure 1**

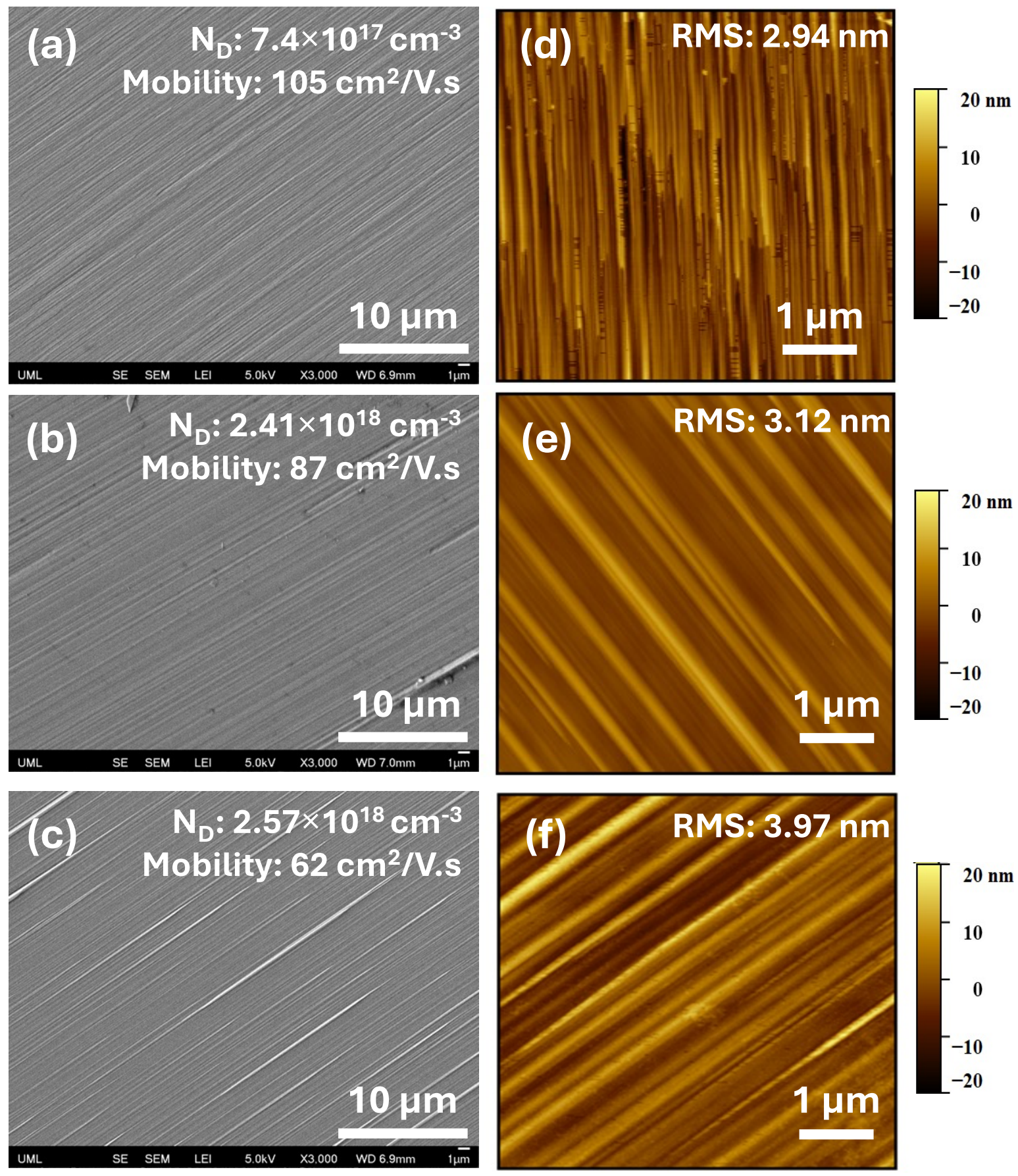

**Figure 1.** Surface morphology of LPCVD-grown Ge-doped β-$Ga_2O_3$ homoepitaxial films on (010) β-$Ga_2O_3$ substrates. (a-c) Surface-view SEM images of films with room-temperature carrier concentrations of $7.4 \times 10^{17}$, $2.41 \times 10^{18}$, and $2.57 \times 10^{18}$ $cm^{-3}$. (d-f) Corresponding AFM images acquired over a $5 \times 5$ $\mu m^2$ scan area, showing elongated terrace-like surface features and RMS roughness values of 2.94, 3.12, and 3.97 nm, respectively.

## Figure 2

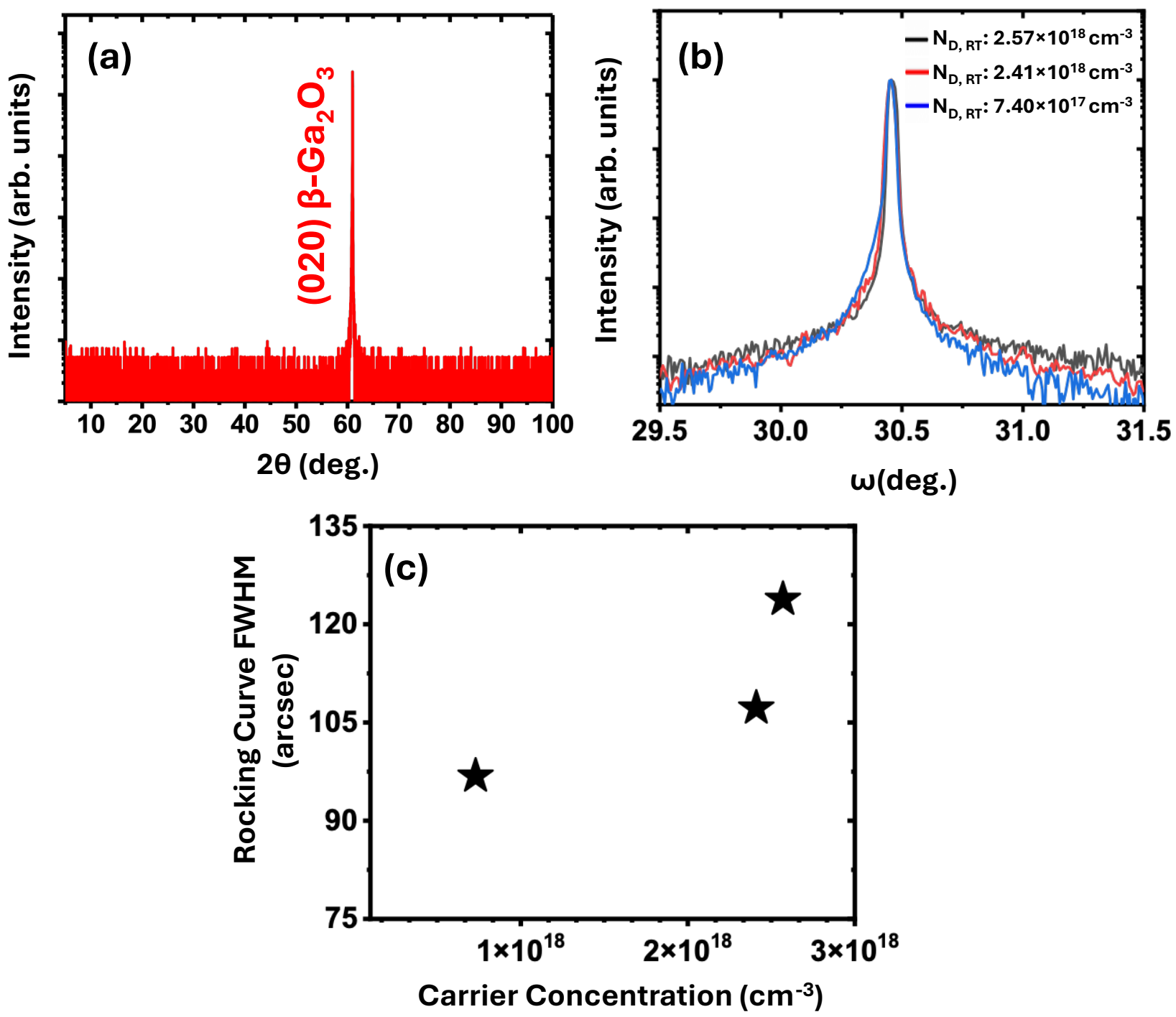

**Figure 2.** Structural characterization of LPCVD-grown Ge-doped β-$Ga_2O_3$ homoepitaxial films. (a) Wide-range XRD $\omega - 2\theta$ scan of the film with a carrier concentration of $7.4 \times 10^{17}$ $\mathrm{cm}^{-3}$. (b) XRD $\omega - rocking$ curves measured from the β-$Ga_2O_3$ (020) reflection for films with carrier concentrations of $7.4 \times 10^{17}$, $2.41 \times 10^{18}$, and $2.57 \times 10^{18}$ $\mathrm{cm}^{-3}$. (c) Variation of the rocking-curve full width at half maximum (FWHM) as a function of carrier concentration, showing a gradual increase in FWHM with increasing Ge concentration.

## Figure 3

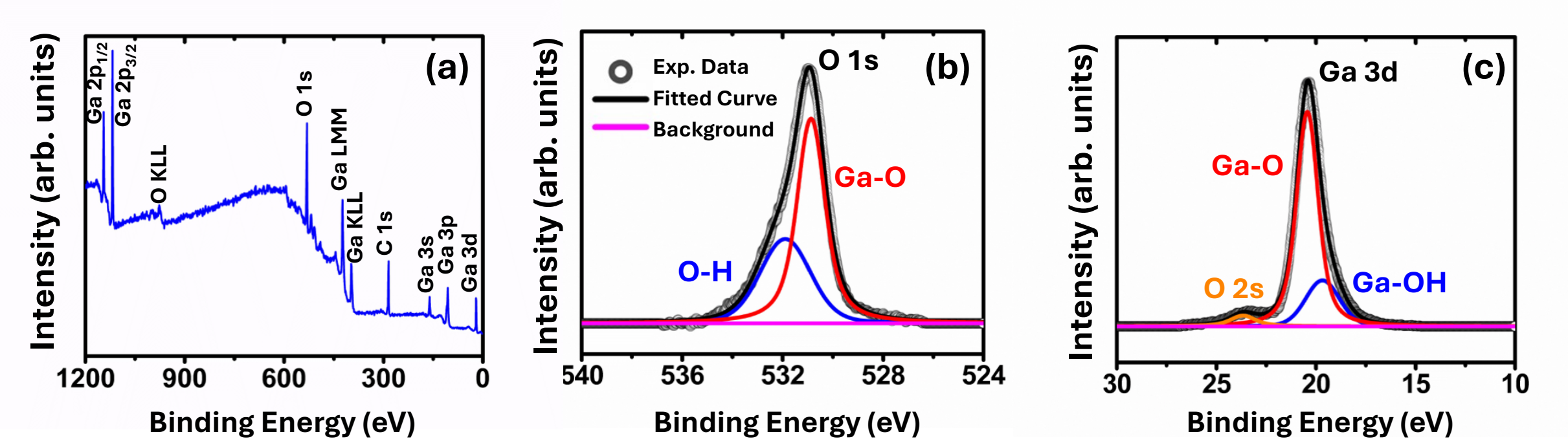

**Figure 3.** XPS analysis of the Ge-doped β-$Ga_2O_3$ homoepitaxial film with a carrier concentration of $7.4 \times 10^{17}$ $\mathrm{cm}^{-3}$. (a) Survey spectrum showing the characteristic Ga and O core-level features of β-$Ga_2O_3$. (b) High-resolution O 1s spectrum deconvoluted into a dominant lattice oxygen (Ga-O) component and a higher-binding-energy contribution associated with surface hydroxyl (O-H) species. (c) High-resolution Ga 3d spectrum consisting primarily of the Ga-O bonding state together with a Ga-OH-related component.

**Figure 4**

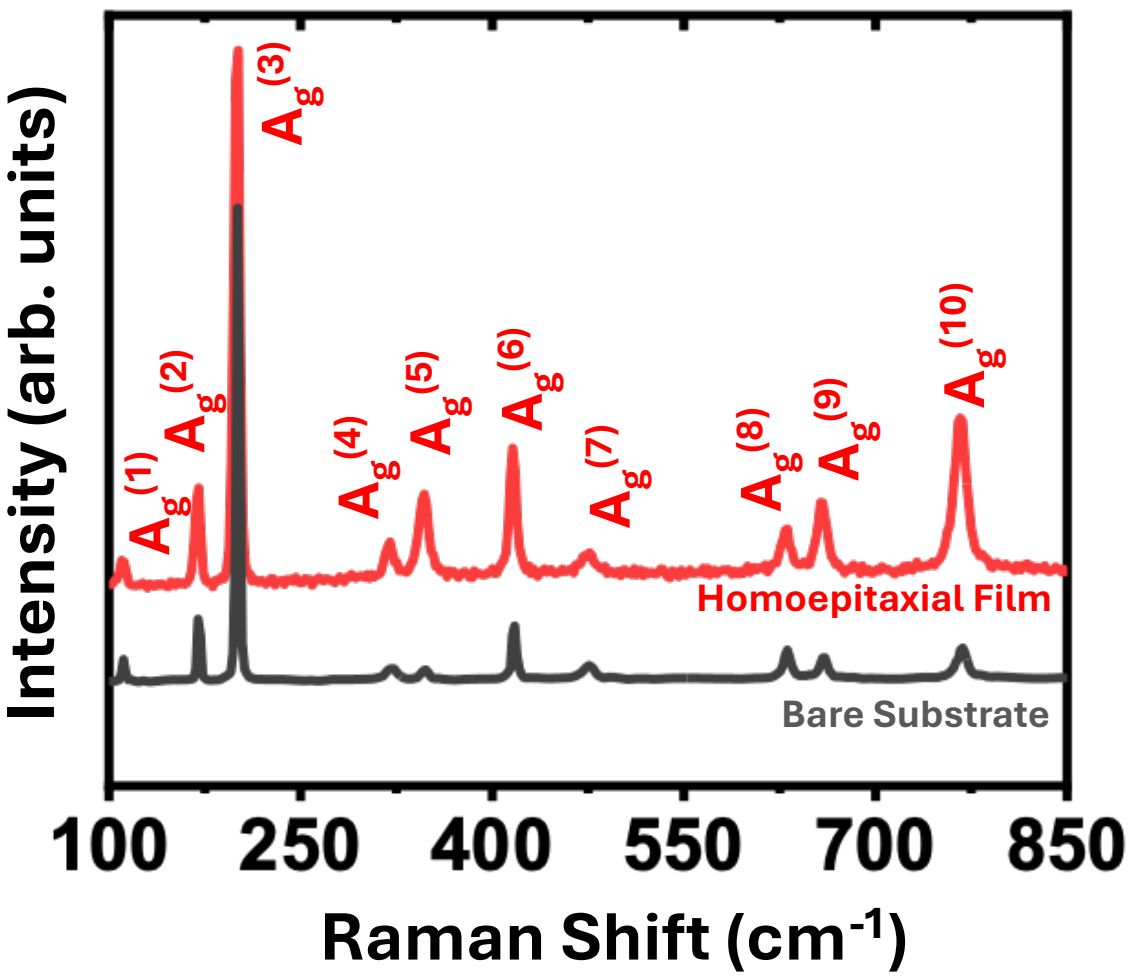

**Figure 4.** Room temperature Raman spectra of the Ge-doped β-$Ga_2O_3$ homoepitaxial film with a carrier concentration of $7.4 \times 10^{17}$ cm$^{-3}$, together with the (010) β-$Ga_2O_3$ bare substrate spectrum.

**Figure 5**

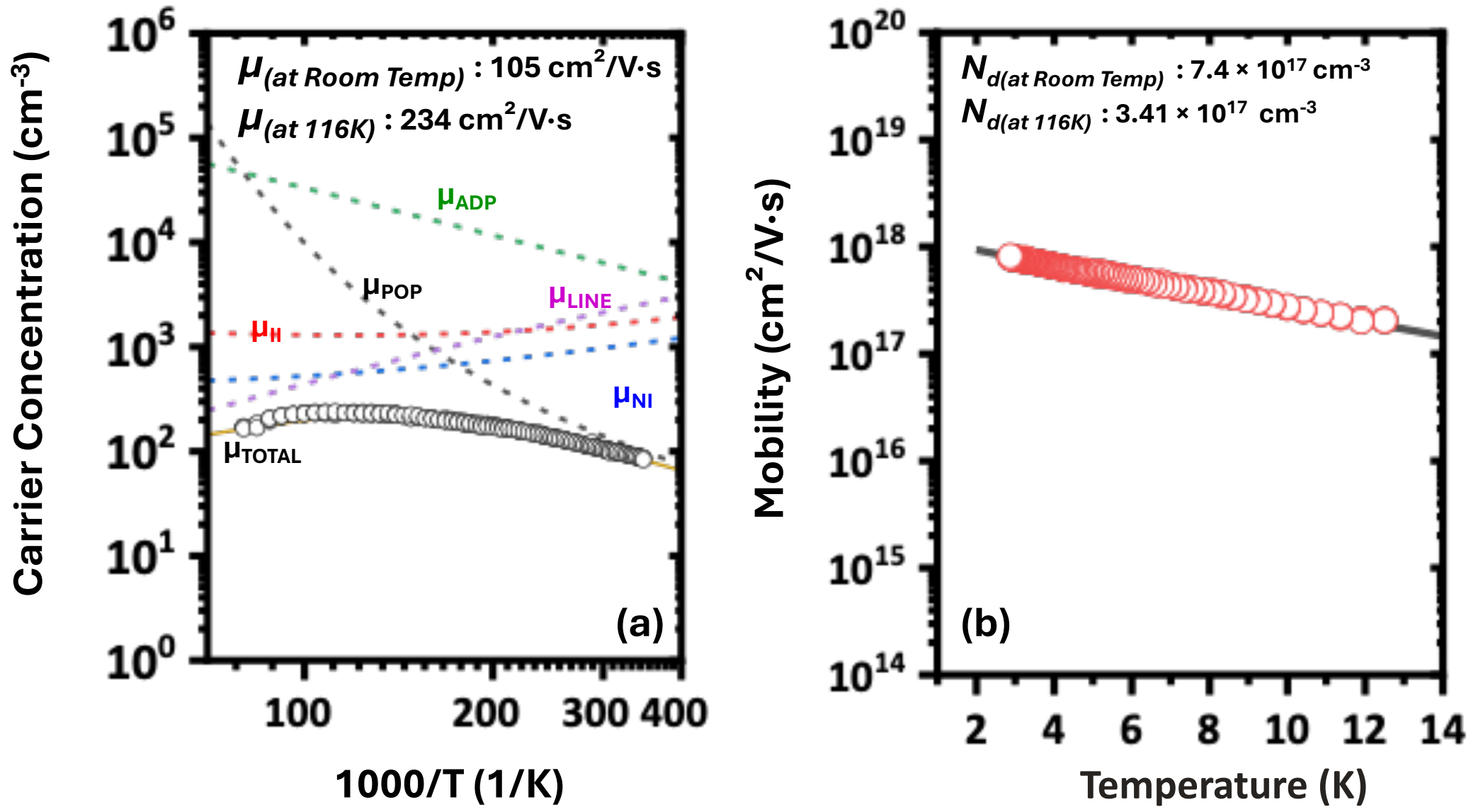

**Figure 5.** Temperature-dependent transport characteristics of the Ge-doped β-$Ga_2O_3$ homoepitaxial film with a room temperature carrier concentration of $7.4 \times 10^{17}$ $cm^{-3}$ and mobility of 105 $cm^2$/V.s. (a) Calculated mobility contributions arising from polar optical phonon scattering ($\mu_{\mathrm{POP}}$), acoustic deformation potential scattering ($\mu_{\mathrm{ADP}}$), ionized impurity scattering ($\mu_{\mathrm{II}}$), neutral impurity scattering ($\mu_{\mathrm{NI}}$), and charged line-defect scattering ($\mu_{\mathrm{line}}$), together with the resulting total modeled mobility ($\mu_{\mathrm{total}}$). Open circles represent the experimentally measured Hall mobility, which reaches a peak value of 234 $cm^2$/V·s at 116 K. (b) Temperature-dependent carrier concentration plotted as a function of 1000/T, showing the experimental Hall data and the corresponding fit obtained using the charge-neutrality model.

**Figure 6**

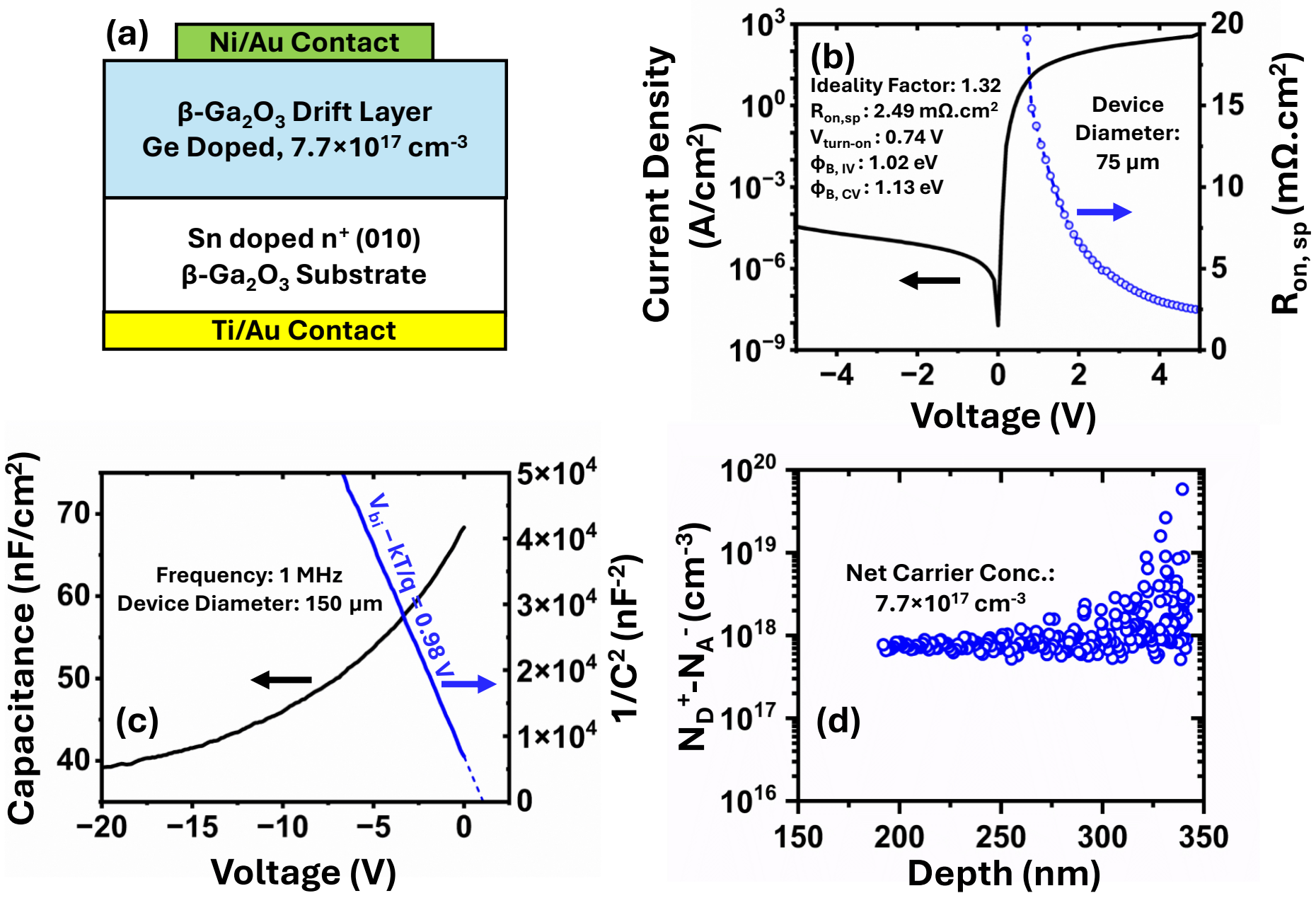

**Figure 6.** Electrical characteristics of the vertical Ni/β-$Ga_2O_3$ Schottky barrier diode fabricated on a 2.54 µm thick Ge-doped β-$Ga_2O_3$ drift layer with a carrier concentration of $7.7 \times 10^{17}$ cm$^{-3}$. (a) Schematic cross-section of the fabricated vertical Schottky diode consisting of a Ni/Au Schottky contact on the Ge-doped β-$Ga_2O_3$ drift layer and a Ti/Au backside Ohmic contact on the conductive Sn-doped (010) β-$Ga_2O_3$ substrate. (b) Room-temperature current density-voltage (J-V) characteristics. (c) Room-temperature capacitance-voltage (C-V) and corresponding $1/C^2 - V$ characteristics measured at 1 MHz. (d) Net carrier concentration profile extracted from the C-V measurements.